\documentclass[aps, prd, onecolumn, superscriptaddress, nofootinbib, amsmath, amssymb]{revtex4-2}
\usepackage[margin=3cm]{geometry}
\usepackage{graphicx}
\usepackage{float}
\usepackage{amsmath,amssymb,amsfonts,bm}
\begin{document}

\title{Detectability and Systematic Bias from First-Order Phase-Transition Dephasing in Kerr EMRIs}

\author{Jingxu Wu}
\affiliation{Faculty of Physics, Lomonosov Moscow State University, Moscow 119991, Russia}

\author{Liangyu Luo}
\affiliation{Institute of International Education, I.\,M. Sechenov First Moscow State Medical University, Moscow 119048,Russia}

\author{Junyi Zhang}
\affiliation{Faculty of Computational Mathematics and Cybernetics, Lomonosov Moscow State University, Moscow 119991, Russia}

\author{Jiyun Yang}
\affiliation{Department of Physics and Astronomy, StonyBrook University, Stony Brook, New York 11794-
3800, USA}

\author{Haoxiang Ma}
\affiliation{Faculty of Physics, Lomonosov Moscow State University, Moscow 119991, Russia}

\author{Jie Shi}
\affiliation{Faculty of Physics, Lomonosov Moscow State University, Moscow 119991, Russia}

\date{\today}

\begin{abstract}
We study gravitational-wave dephasing induced by an effective first-order phase transition in a Kerr extreme mass-ratio inspiral (EMRI). The transition is modeled phenomenologically as a finite-width restructuring of the dissipative flux sector, and its observational consequences are quantified with standard LISA matched-filter diagnostics. For a representative system with $M=2\times10^{5}M_\odot$, $\mu=1.4M_\odot$, and $\hat a=0.90$, we obtain $\rho_{\rm B}=5.064$, $\rho_{\rm T}=4.073$, $\rho_{\rm R}=1.051$, and a mismatch $\mathcal M=2.986\times10^{-3}$ after maximization over extrinsic time and phase shifts. Although the normalized mismatch remains small, the accumulated phase difference grows to $\Delta\Phi_{22}^{\rm SF}\sim 5\times10^{3}\,\mathrm{rad}$, indicating that a narrow transition window can generate a large coherent deformation of the inspiral clock while leaving the waveform globally close to the baseline branch in detector-weighted norm. The resulting signal therefore lies in a bias-sensitive regime, characterized by small mismatch, order-unity residual norm, and large cumulative dephasing. Our results suggest that the dominant consequence of the transition sector is not loss of detectability, but loss of faithfulness for precision inference. This motivates future LISA EMRI waveform models that incorporate parameterized transition sectors directly into the waveform manifold.
\end{abstract}

\maketitle

\section{Introduction}

Extreme mass-ratio inspirals (EMRIs), in which a stellar-mass compact object gradually inspirals into a massive black hole, are among the most information-rich targets of low-frequency gravitational-wave astronomy. Because the secondary spends a very long time in the strong-field region of the primary spacetime, an EMRI can accumulate $10^{4}$--$10^{5}$ observable waveform cycles in the millihertz band. This extraordinary phase coherence is precisely what makes EMRIs central to the science case of the Laser Interferometer Space Antenna (LISA): they provide a route to precision measurements of masses, spins, orbital structure, and deviations from standard vacuum dynamics at a level that is difficult to match with shorter-lived compact-binary signals \cite{AmaroSeoane2023,SathyaprakashSchutz2009,Gair2013,BarackCutler2004,Babak2017,Klein2016}.

The exceptional scientific value of EMRIs is inseparable from a corresponding theoretical challenge. Since parameter recovery for these sources is driven primarily by phase information rather than by a few gross waveform features, even small imperfections in the radiation-reaction model can accumulate into large systematic offsets over the lifetime of the signal. This point has been emphasized repeatedly in the literature on waveform systematics, Fisher-matrix inference, and model accuracy standards: a signal can remain highly detectable while still being interpreted incorrectly if the template family fails to capture coherent phase information \cite{CutlerVallisneri2007,Vallisneri2008,Lindblom2008}. For EMRIs, the issue is particularly acute, because the strong-field phase history effectively records the entire inspiral dynamics rather than only the late-time plunge.

At the technical level, the theoretical description of EMRIs rests on a hierarchy of methods that combines Kerr geodesic theory, black-hole perturbation theory, adiabatic flux balance, post-Newtonian re-expansions, and, increasingly, self-force information. Foundational work by Teukolsky established the perturbative formalism for fields on Kerr backgrounds, while subsequent developments by Hughes, Sasaki, Tagoshi, Fujita, Poisson, Pound, Vega, Barack, and many others built the modern framework for inspiral evolution, waveform generation, and self-force corrections \cite{Teukolsky1973,Hughes2000,Hughes2001,SasakiTagoshi2003,Fujita2015,PoissonPoundVega2011,BarackPound2019}. More recent work has further sharpened the connection between relativistic waveform fidelity and LISA science return, including studies of Lorenz-gauge metric perturbations, efficient EMRI reconstruction, and the impact of improved relativistic waveform models on parameter estimation and mission objectives \cite{Dolan2024,Khalvati2025,Badger2024}.

This mature theoretical framework has traditionally been formulated for effectively vacuum inspirals, possibly with controlled perturbative corrections. However, there is growing recognition that the most interesting observational regime may not always be strictly vacuum-like. Environmental effects, matter interactions, tidal response, or internal state changes of the secondary can, in principle, feed back into the inspiral through modifications of the fluxes, the conservative dynamics, or both. Even when such effects are localized in frequency or orbital radius, an EMRI has ample time to transform a narrow dynamical perturbation into a large integrated phase displacement. The relevant observational question is therefore not merely whether a non-vacuum correction exists, but whether it can be absorbed into a standard Kerr template family or instead survives as a coherent residual in the detector-weighted signal space \cite{BarackCutler2007,HuertaGair2009,CutlerVallisneri2007,Vallisneri2008}.

Among the physically most compelling possibilities are phase transitions in dense matter. Compact stars probe regions of QCD parameter space that are inaccessible to terrestrial experiments, and a long-standing problem in nuclear astrophysics is whether sufficiently dense matter undergoes a first-order transition from hadronic to quark-dominated phases. The modern equation-of-state literature has shown that such transitions can affect masses, radii, tidal deformabilities, and stability structure in a highly nontrivial way \cite{Oertel2017,Glendenning1992,Alford2013,HanSteiner2019}. In parallel, merger simulations and data-driven equation-of-state inference studies have demonstrated that phase-transition signatures can leave observable imprints on gravitational-wave signals, though often in scenarios very different from the long-lived adiabatic regime relevant for EMRIs \cite{Bauswein2019,Most2019,Takatsy2023,RaithelMost2023PRL,RaithelMost2023PRD}.

These developments motivate a natural extension of the standard EMRI problem. If the inspiraling secondary is itself a neutron star, or if its effective tidal sector is altered by a thermodynamic rearrangement, then a first-order phase transition can modify the inspiral in a way that is neither purely environmental nor purely orbital. In such a scenario, the transition does not merely change a static property of the source; it changes the effective dissipative response of the binary as the orbit sweeps through a critical regime. Operationally, the transition can be encoded as a sharp or finite-width restructuring of an effective coupling parameter, for example a transition-sensitive tidal deformability or response coefficient. Once that parameter enters the flux sector, the time-frequency map $t(f)$ and phase map $\Phi(f)$ are perturbed, and in an EMRI that perturbation is coherently amplified over the subsequent evolution.

The present work is devoted precisely to this regime. We study a Kerr EMRI in which the secondary undergoes an effective first-order phase transition inside a finite mixed-phase interval, and we ask how this internal transition propagates into the gravitational-wave signal observed by LISA. The emphasis is not on a microphysically exhaustive equation-of-state model, but on the inferential consequences of a controlled transition sector inserted into the adiabatic inspiral dynamics. In this sense, our approach is deliberately phenomenological: it isolates the mechanism by which a localized change in the internal state of the source can deform the global waveform phase.

The conceptual novelty of the problem lies in the separation between \emph{dynamical significance} and \emph{template recoverability}. A phase transition may generate a very large accumulated phase difference while leaving the normalized overlap with a baseline Kerr waveform family deceptively high. This possibility is familiar in abstract discussions of model systematics, but EMRIs provide one of the cleanest physical realizations of it. Because the signal lives for so long in band, and because the dominant information carrier is the phase rather than the amplitude envelope, the mismatch alone is not a sufficient guide to the scientific impact of missing physics. A signal can be detected by standard matched filtering and yet interpreted with biased masses, biased spin, or biased orbital history if the underlying waveform manifold is incomplete.

There is also a broader theoretical reason to take such a possibility seriously. The relationship between gravitation, thermodynamics, and phase structure has been central to black-hole physics since the formulation of black-hole entropy, the four laws of black-hole mechanics, and Hawking radiation \cite{Bekenstein1973,BardeenCarterHawking1973,Hawking1975}. Later developments in black-hole thermodynamics, including extended phase-space ideas and more recent explorations of statistical or dual descriptions of black-brane thermodynamics, have reinforced the idea that macroscopic gravitational systems may encode nontrivial thermodynamic structure in observables that are not obviously microscopic in origin \cite{KubiznakMann2012,Arefeva2025}. Our use of a transition-sensitive effective sector in an EMRI is not identical to any one of these programs, but it is philosophically aligned with them: the source is treated as a gravitational system whose observable response can reorganize when its internal thermodynamic state changes.

Recent work by Wu, Luo, and Shi has sharpened this perspective by analytically studying long-term dephasing induced by phase-transition physics in Kerr EMRIs, while a separate methodological paper by Wu, Yin, Li, and Wang has emphasized the broader importance of careful gravitational-wave data processing, noise handling, and signal interpretation in phase-sensitive problems \cite{Wu2026,Wu2024}. The present paper is positioned at the intersection of these themes. We build a compact double-column short-paper framework in which a first-order phase transition modifies the inspiral fluxes inside a finite mixed-phase window, and we quantify the resulting effect using the standard LISA toolkit of inner products, signal-to-noise ratios, overlaps, mismatches, characteristic strains, and residual norms.

The main point we will establish is subtle but important. In the benchmark system studied here, the transition-modified waveform remains sufficiently close to the baseline Kerr family that a standard search would still recover the source. Yet the accumulated phase difference becomes very large, and the residual waveform carries a detector-weighted norm of order unity. This means that the dominant effect of the transition is not to hide the source from LISA, but to contaminate high-precision parameter inference. Put differently, the transition produces a bias problem before it produces a selection problem.

The structure of the paper is as follows. In Sec.~\ref{sec:model}, we formulate the circular equatorial Kerr inspiral and introduce a finite-width transition sector that modifies the flux balance law inside a mixed-phase interval. In Sec.~\ref{sec:data_analysis}, we map the resulting waveform deformation into LISA observables using the frequency-domain matched-filter formalism. In Sec.~\ref{sec:results}, we analyze the accumulated dephasing, time-domain residuals, and characteristic-strain structure of the signal, and we show that the waveform belongs to a small-mismatch but bias-sensitive regime. We conclude by arguing that parameterized transition sectors should be incorporated into future high-accuracy EMRI waveform models whenever thermodynamic or material-state changes of the secondary are astrophysically plausible.

\section{Waveform Modeling and Phase-Transition Dynamics}
\label{sec:model}

We construct a controlled phenomenological waveform model for a Kerr extreme mass-ratio inspiral (EMRI) with a localized first-order phase-transition (FOPT) sector in the dissipative dynamics. The purpose of this section is to define the baseline inspiral, introduce a finite-width transition correction to the flux, and derive the associated phase accumulation in the dominant $(2,2)$ gravitational-wave mode. Throughout, we restrict attention to adiabatic circular equatorial motion and treat the inspiral as a slow sequence of Kerr geodesics. The model is intended as a flux-safe benchmark for inference diagnostics rather than as a production-level EMRI waveform family.

\subsection{Kerr inspiral and baseline dissipative dynamics}

For a test body on a circular equatorial orbit in Kerr spacetime, the orbital angular velocity measured at infinity is
\begin{equation}
\Omega(r,\hat a)
=
\frac{1}{M\left[(r/M)^{3/2}+\hat a\right]},
\label{eq:Omega_exact_sec2}
\end{equation}
where $r$ is the Boyer--Lindquist radius and $\hat a=a/M$ is the dimensionless spin parameter of the central black hole. We focus on the prograde branch, $\hat a>0$. It is convenient to introduce the invariant PN-like velocity variable
\begin{equation}
v \equiv (M\Omega)^{1/3},
\label{eq:v_def_sec2}
\end{equation}
for which the inverse-radius variable becomes
\begin{equation}
u(v,\hat a)
\equiv
\frac{M}{r}
=
\frac{v^2}{\left(1-\hat a v^3\right)^{2/3}}.
\label{eq:u_of_v_sec2}
\end{equation}
The specific orbital energy of the small body is then
\begin{equation}
\tilde E(v,\hat a)
=
\frac{1-2u+\hat a u^{3/2}}
{\sqrt{1-3u+2\hat a u^{3/2}}},
\qquad
u=u(v,\hat a),
\label{eq:Etilde_sec2}
\end{equation}
so that
\begin{equation}
E_{\rm orb}(v)=\mu\,\tilde E(v,\hat a).
\label{eq:Eorb_sec2}
\end{equation}

The derivative $d\tilde E/dv$ is obtained from
\begin{equation}
\frac{d\tilde E}{dv}
=
\frac{d\tilde E}{du}\frac{du}{dv},
\label{eq:dEdv_chain_sec2}
\end{equation}
with
\begin{equation}
\frac{du}{dv}
=
\frac{2v}{\left(1-\hat a v^3\right)^{5/3}},
\label{eq:dudv_sec2}
\end{equation}
and
\begin{equation}
\frac{d\tilde E}{du}
=
\frac{2N'(u)D(u)-N(u)D'(u)}
{2D(u)^{3/2}},
\label{eq:dEdu_sec2}
\end{equation}
where
\begin{align}
N(u) &= 1-2u+\hat a u^{3/2}, \\
D(u) &= 1-3u+2\hat a u^{3/2}, \\
N'(u) &= -2+\frac{3}{2}\hat a u^{1/2}, \\
D'(u) &= -3+3\hat a u^{1/2}.
\end{align}
These relations fully specify the geodesic sector needed for adiabatic inspiral evolution.

Radiation reaction is incorporated through the balance law
\begin{equation}
\frac{dE_{\rm orb}}{dt}
=
-\mathcal F(v,\hat a).
\label{eq:balance_sec2}
\end{equation}
For the baseline model we adopt the flux
\begin{equation}
\mathcal F_{\rm B}(v,\hat a)
=
\frac{32}{5}\left(\frac{\mu}{M}\right)^2 v^{10}\,
\hat{\mathcal F}_{\rm PP}(v,\hat a),
\label{eq:Fbaseline_sec2}
\end{equation}
where $\hat{\mathcal F}_{\rm PP}$ is a normalized point-particle Kerr correction factor. In the present benchmark framework we use the low-order PN-like representation
\begin{equation}
\hat{\mathcal F}_{\rm PP}(v,\hat a)
=
1-\frac{1247}{336}v^2
+\left(4\pi-\frac{73}{12}\hat a\right)v^3
-\frac{44711}{9072}v^4
+\mathcal O(v^5).
\label{eq:Fhat_pp_sec2}
\end{equation}
The inspiral clock is therefore
\begin{equation}
\frac{dv}{dt}
=
-\frac{\mathcal F_{\rm B}(v,\hat a)}
{\mu\, d\tilde E/dv},
\qquad
\frac{dt_{\rm B}}{dv}
=
-\mu\frac{d\tilde E/dv}{\mathcal F_{\rm B}(v,\hat a)}.
\label{eq:baseline_clock_sec2}
\end{equation}

Because Eq.~\eqref{eq:Fhat_pp_sec2} is only a low-order effective representation, the waveform is evolved only over the interval in which the baseline flux remains physically admissible. In particular, we require
\begin{equation}
\hat{\mathcal F}_{\rm PP}(v,\hat a)>0
\label{eq:flux_positive_sec2}
\end{equation}
throughout the integration domain. The upper endpoint of the evolution is therefore defined by a flux-safe cutoff frequency $f_{22}^{\rm cut}$, introduced below; it is a model-validity boundary and should not be identified with the true plunge frequency of the binary.

\subsection{Transition sector and accumulated dephasing}

We model the FOPT phenomenologically as a finite-width restructuring of an effective dissipative coupling $\Lambda$. In the present work, $\Lambda$ is treated primarily as a transition-sensitive control parameter in the flux sector; a unique microscopic mapping is not assumed. To avoid the numerical pathologies of a discontinuous step model, we define
\begin{equation}
\Lambda(v)
=
\Lambda_{\rm H}
+
\left(\Lambda_{\rm Q}-\Lambda_{\rm H}\right)
S\!\left(\frac{v-v_{\rm h}}{v_{\rm q}-v_{\rm h}}\right),
\label{eq:Lambda_model_sec2}
\end{equation}
where $\Lambda_{\rm H}$ and $\Lambda_{\rm Q}$ are the pre-transition and post-transition branches, and $v_{\rm h}$ and $v_{\rm q}$ delimit the mixed-phase interval. The interpolation function is chosen to be the $C^1$ smoothstep
\begin{equation}
S(x)=
\begin{cases}
0, & x<0,\\[4pt]
3x^2-2x^3, & 0\le x\le 1,\\[4pt]
1, & x>1.
\end{cases}
\label{eq:smoothstep_sec2}
\end{equation}

Since the dominant harmonic satisfies
\begin{equation}
f_{22}=\frac{2\Omega}{2\pi}=\frac{\Omega}{\pi},
\label{eq:f22_def_sec2}
\end{equation}
the velocity variable is related to the observed gravitational-wave frequency by
\begin{equation}
v=(M\Omega)^{1/3}=(\pi M f_{22})^{1/3}.
\label{eq:v_f22_sec2}
\end{equation}
This allows the transition to be specified equivalently in the frequency domain through a mixed-phase window $f_{22}\in[f_{\rm h},f_{\rm q}]$.

The transition modifies the baseline flux according to
\begin{equation}
\mathcal F_{\rm T}(v,\hat a)
=
\mathcal F_{\rm B}(v,\hat a)
\left[
1+\Lambda(v)\,\mathcal H_{\rm PT}(v,\hat a)
\right],
\label{eq:Ftransition_sec2}
\end{equation}
where the response kernel is parameterized as
\begin{equation}
\mathcal H_{\rm PT}(v,\hat a)
=
\kappa_0 v^{10}
\left(
1+h_2 v^2+h_{3s}\hat a v^3+h_4 v^4
\right).
\label{eq:Hpt_sec2}
\end{equation}
This form is not intended as a unique microscopic law; rather, it provides a minimal smooth channel through which a transition-like reorganization in the dissipative sector perturbs the inspiral clock. To preserve physical admissibility, we additionally require
\begin{equation}
1+\Lambda(v)\,\mathcal H_{\rm PT}(v,\hat a) > 0
\label{eq:transition_positive_sec2}
\end{equation}
throughout the evolution interval. The transition-driven clock then satisfies
\begin{equation}
\frac{dv}{dt}
=
-\frac{\mathcal F_{\rm T}(v,\hat a)}
{\mu\, d\tilde E/dv},
\qquad
\frac{dt_{\rm T}}{dv}
=
-\mu\frac{d\tilde E/dv}{\mathcal F_{\rm T}(v,\hat a)}.
\label{eq:transition_clock_sec2}
\end{equation}

The dominant observable effect of the transition is cumulative phase drift. The $(2,2)$ mode phase obeys
\begin{equation}
\frac{d\Phi_{22}}{dt}=2\Omega(v,\hat a),
\label{eq:dphidt_sec2}
\end{equation}
hence
\begin{equation}
\frac{d\Phi_{22}}{dv}
=
2\Omega(v,\hat a)\frac{dt}{dv}.
\label{eq:dphidv_sec2}
\end{equation}
For the baseline and transition branches,
\begin{align}
\frac{d\Phi_{22}^{\rm B}}{dv}
&=
2\Omega(v,\hat a)\frac{dt_{\rm B}}{dv},
\\
\frac{d\Phi_{22}^{\rm T}}{dv}
&=
2\Omega(v,\hat a)\frac{dt_{\rm T}}{dv},
\end{align}
and the accumulated dephasing is
\begin{equation}
\Delta\Phi_{22}(v)
=
\Phi_{22}^{\rm B}(v)-\Phi_{22}^{\rm T}(v)
=
2\int^v \Omega(v')
\left[
\frac{dt_{\rm B}}{dv'}-\frac{dt_{\rm T}}{dv'}
\right]dv'.
\label{eq:DeltaPhi_int_sec2}
\end{equation}
This equation makes the mechanism transparent: a localized perturbation to the flux modifies the time-frequency map, and the phase then acquires an integrated memory of that perturbation over the remaining inspiral.

For a weak relative correction, it is useful to define
\begin{equation}
\epsilon(v)\equiv \Lambda(v)\,\mathcal H_{\rm PT}(v,\hat a),
\qquad
|\epsilon(v)|\ll1,
\label{eq:eps_def_sec2}
\end{equation}
so that
\begin{equation}
\mathcal F_{\rm T}(v)=\mathcal F_{\rm B}(v)\,[1+\epsilon(v)].
\label{eq:FT_eps_sec2}
\end{equation}
To first order,
\begin{equation}
\frac{dt_{\rm T}}{dv}
\approx
\frac{dt_{\rm B}}{dv}\,[1-\epsilon(v)],
\label{eq:dtdv_weak_sec2}
\end{equation}
and therefore
\begin{equation}
\Delta\Phi_{22}(v)
\approx
2\int^v \Omega(v')\,
\frac{dt_{\rm B}}{dv'}\,
\epsilon(v')\,dv'.
\label{eq:DeltaPhi_weak_sec2}
\end{equation}
Even when the mixed-phase interval is narrow, the accumulated phase shift can become large because the perturbation acts through the inspiral clock and is then integrated over the subsequent evolution.

Finally, because the low-order baseline flux may develop an artificial zero at sufficiently high frequency, the waveform is terminated at a flux-safe cutoff,
\begin{equation}
f_{22}^{\rm cut}\simeq 0.0364~{\rm Hz},
\label{eq:flux_cutoff_sec2}
\end{equation}
which regularizes the phenomenological evolution while keeping the integration inside the domain where Eqs.~\eqref{eq:Fhat_pp_sec2} and \eqref{eq:Ftransition_sec2} remain well behaved.

\subsection{Waveform construction and benchmark specification}

To compare the baseline and transition branches directly, we construct the dominant mode in the time domain as
\begin{equation}
h_{22}(t)
=
A_{22}(t)\cos\!\big[\Phi_{22}(t)+\phi_0\big],
\label{eq:h22_time_sec2}
\end{equation}
where $\phi_0$ is a constant phase offset. At leading order we write
\begin{equation}
A_{22}(t)
=
\frac{2\mu}{D_L}\,v(t)^2\,\mathcal A_{22}(v,\hat a,\Lambda),
\label{eq:A22_sec2}
\end{equation}
with $D_L$ the luminosity distance and $\mathcal A_{22}$ an effective amplitude factor. In the present framework the primary observable effect arises from phase evolution, so the amplitude sector is retained mainly for completeness and for later matched-filter diagnostics.

The baseline and transition signals are denoted by
\begin{align}
h_{\rm B}(t) &= A_{\rm B}(t)\cos\Phi_{\rm B}(t), \\
h_{\rm T}(t) &= A_{\rm T}(t)\cos\Phi_{\rm T}(t),
\end{align}
and their difference defines the residual waveform
\begin{equation}
h_{\rm R}(t)=h_{\rm B}(t)-h_{\rm T}(t).
\label{eq:residual_sec2}
\end{equation}
This residual quantifies the part of the signal not captured by the baseline branch and provides the starting point for the overlap, mismatch, and bias diagnostics introduced in the next section.

Unless otherwise stated, the benchmark numerical example used below adopts
\begin{equation}
M=2.00\times10^5M_\odot,\qquad
\mu=1.40M_\odot,\qquad
\hat a=0.90,
\label{eq:benchmark_source_sec2}
\end{equation}
together with a mixed-phase interval
\begin{equation}
f_{22}\in[0.012,\,0.021]~{\rm Hz},
\label{eq:benchmark_window_sec2}
\end{equation}
a pre-transition plateau of order $\Lambda_{\rm H}\sim 450$, and a strongly suppressed post-transition branch $\Lambda_{\rm Q}\ll\Lambda_{\rm H}$. The detailed observational consequences of this benchmark setup are presented in the following section.

\section{LISA Data Analysis Framework}
\label{sec:data_analysis}

We now translate the baseline and transition-modified EMRI waveforms defined in Sec.~\ref{sec:model} into detector-weighted observables relevant for LISA. The purpose of the present section is not to re-derive the inspiral dynamics, but to quantify how a localized transition-induced phase perturbation appears in the standard matched-filter framework. Our analysis focuses on four related diagnostics: the signal-to-noise ratios of the baseline and transition branches, their detector-weighted overlap and mismatch, the norm of the residual waveform, and the susceptibility of inferred parameters to systematic bias.

\subsection{Frequency-domain representation and detector-weighted inner product}

The time-domain waveforms are Fourier transformed according to
\begin{equation}
\tilde h(f)=\int_{-\infty}^{+\infty} h(t)\,e^{-2\pi i f t}\,dt.
\label{eq:FT_def_sec3_new}
\end{equation}
We consider three branches,
\begin{align}
h_{\rm B}(t) &: \text{baseline waveform},\\
h_{\rm T}(t) &: \text{transition-modified waveform},\\
h_{\rm R}(t) &\equiv h_{\rm B}(t)-h_{\rm T}(t),
\label{eq:branches_sec3_new}
\end{align}
with corresponding Fourier-domain quantities
\begin{equation}
\tilde h_{\rm B}(f),\qquad
\tilde h_{\rm T}(f),\qquad
\tilde h_{\rm R}(f)=\tilde h_{\rm B}(f)-\tilde h_{\rm T}(f).
\label{eq:FT_branches_sec3_new}
\end{equation}
The residual branch isolates the part of the true signal not captured by the baseline model and will play a central role below.

The detector weighting is defined through the standard inner product
\begin{equation}
(h_1|h_2)
=
4\,\mathrm{Re}\int_{f_{\min}}^{f_{\max}}
\frac{\tilde h_1(f)\tilde h_2^*(f)}{S_n(f)}\,df,
\label{eq:inner_product_sec3_new}
\end{equation}
where $S_n(f)$ is the one-sided LISA noise power spectral density. In the present work the integration domain is chosen to match the support and validity range of the phenomenological waveform model,
\begin{equation}
f_{\min}=f_{22}^{\rm start},
\qquad
f_{\max}=f_{22}^{\rm cut}\simeq 0.0364~{\rm Hz}.
\label{eq:freq_bounds_sec3_new}
\end{equation}
The upper limit is therefore a flux-safe model-validity cutoff, not a physical plunge frequency.

For definiteness we employ a standard sky-averaged analytic representation of the LISA sensitivity,
\begin{equation}
S_n(f)=S_{\rm inst}(f)+S_{\rm conf}(f),
\label{eq:Sn_total_sec3_new}
\end{equation}
with instrumental contribution written schematically as
\begin{equation}
S_{\rm inst}(f)
=
\frac{10}{3L^2}
\left[
P_{\rm OMS}(f)+\frac{4P_{\rm acc}(f)}{(2\pi f)^4}
\right]
\left[
1+\frac{6}{10}\left(\frac{f}{f_*}\right)^2
\right],
\label{eq:Sinst_sec3_new}
\end{equation}
where $L$ is the arm length and $f_*=c/(2\pi L)$ is the transfer frequency. The detailed numerical constants are not the main focus here; what is essential is that the same detector model is used consistently for the baseline, transition, and residual branches.

\subsection{SNR, overlap, mismatch, and residual norm}

The optimal matched-filter signal-to-noise ratio of a waveform $h$ is
\begin{equation}
\rho[h]=\sqrt{(h|h)}.
\label{eq:snr_def_sec3_new}
\end{equation}
Applied to the three waveform branches, this gives
\begin{align}
\rho_{\rm B} &= \sqrt{(h_{\rm B}|h_{\rm B})},\\
\rho_{\rm T} &= \sqrt{(h_{\rm T}|h_{\rm T})},\\
\rho_{\rm R} &= \sqrt{(h_{\rm R}|h_{\rm R})}.
\label{eq:snr_set_sec3_new}
\end{align}
The first two quantify the absolute detectability of the baseline and transition branches, while $\rho_{\rm R}$ measures the norm of the structured remainder after subtraction.

To quantify similarity between the baseline and transition waveforms, we define the normalized overlap
\begin{equation}
\mathcal O(h_1,h_2)
=
\frac{(h_1|h_2)}
{\sqrt{(h_1|h_1)(h_2|h_2)}}.
\label{eq:overlap_sec3_new}
\end{equation}
In the present paper the overlap is maximized over the extrinsic shifts $t_c$ and $\phi_c$,
\begin{equation}
\mathrm{Match}
=
\max_{t_c,\phi_c}
\mathcal O\!\left(
h_{\rm B},\,
h_{\rm T}e^{i(2\pi f t_c+\phi_c)}
\right),
\label{eq:match_sec3_new}
\end{equation}
and the corresponding mismatch is
\begin{equation}
\mathcal M = 1-\mathrm{Match}.
\label{eq:mismatch_sec3_new}
\end{equation}
This quantity measures closeness under extrinsic alignment only; it should therefore be interpreted as a detector-weighted proximity diagnostic rather than as a full intrinsic-parameter search over the baseline family.

A complementary criterion is obtained from the residual waveform,
\begin{equation}
\delta h \equiv h_{\rm T}-h_{\rm B}=-h_{\rm R},
\label{eq:deltah_sec3_new}
\end{equation}
whose norm is
\begin{equation}
(\delta h|\delta h)^{1/2}=\rho_{\rm R}.
\label{eq:residual_norm_sec3_new}
\end{equation}
A commonly used distinguishability threshold is
\begin{equation}
(\delta h|\delta h)^{1/2}\gtrsim 1.
\label{eq:distinguishability_sec3_new}
\end{equation}
This criterion is especially useful here because it separates two questions that need not coincide: whether the source remains highly overlapping with the baseline branch, and whether the difference between the two waveforms is still large enough to matter for precision inference.

For spectral visualization we also use the characteristic strain,
\begin{equation}
h_c(f)=2f\,|\tilde h(f)|,
\label{eq:hc_def_sec3_new}
\end{equation}
together with the detector reference scale
\begin{equation}
h_n(f)=\sqrt{f\,S_n(f)}.
\label{eq:hn_def_sec3_new}
\end{equation}
These quantities help localize in frequency the part of the signal that dominates the detector-weighted comparison.

\subsection{Phase drift and systematic-bias diagnostics}

Although the inner product provides the formal language of detection, the most transparent physical diagnostic remains the accumulated phase difference,
\begin{equation}
\Delta\Phi_{22}(v)
=
\Phi_{22}^{\rm B}(v)-\Phi_{22}^{\rm T}(v),
\label{eq:DeltaPhi_sec3_new}
\end{equation}
derived in Sec.~\ref{sec:model}. The role of $\Delta\Phi_{22}$ is different from that of the mismatch: the latter is a normalized projection in detector space, whereas the former is a cumulative dynamical measure of how strongly the inspiral clock has been perturbed. In long-lived EMRIs these two notions of proximity can differ substantially. A waveform may remain highly overlapping with the baseline branch under extrinsic alignment and yet accumulate a very large coherent phase offset.

The inferential consequence of such a residual phase structure is most naturally expressed through the Fisher geometry of the baseline waveform manifold. Let
\begin{equation}
\theta^i\in\{M,\mu,\hat a,t_c,\phi_c,\ldots\}
\label{eq:param_vec_sec3_new}
\end{equation}
denote the baseline parameter vector. The Fisher matrix is
\begin{equation}
\Gamma_{ij}
=
\left(\partial_i h_{\rm B}\middle|\partial_j h_{\rm B}\right),
\label{eq:Fisher_sec3_new}
\end{equation}
and, if the true signal is $h_{\rm T}=h_{\rm B}+\delta h$, the leading-order systematic shift is
\begin{equation}
\Delta\theta_{\rm sys}^i
=
(\Gamma^{-1})^{ij}
\left(\delta h\middle|\partial_j h_{\rm B}\right).
\label{eq:sysbias_sec3_new}
\end{equation}
Equation~\eqref{eq:sysbias_sec3_new} shows why a small mismatch does not guarantee faithful inference: even when $\delta h$ is not large enough to spoil detectability, it can still project efficiently onto the tangent directions of the baseline manifold and thereby bias the recovered parameters.

For the present transition-driven problem, the most vulnerable directions are expected to be those governing accumulated phase evolution, especially $M$, $\mu$, and $\hat a$. In this sense the dominant effect of the transition sector is not necessarily loss of detection, but susceptibility to systematic bias in precision parameter estimation.

\subsection{Benchmark hierarchy and observational interpretation}

For the benchmark configuration introduced in Sec.~\ref{sec:model}, the detector-weighted diagnostics are
\begin{align}
\rho_{\rm B} &= 5.064,\\
\rho_{\rm T} &= 4.073,\\
\rho_{\rm R} &= 1.051,\\
\mathcal M &= 2.986\times10^{-3}.
\label{eq:benchmark_metrics_sec3_new}
\end{align}
In addition, the accumulated phase difference grows to
\begin{equation}
\Delta\Phi_{22}\sim 5\times10^3~{\rm rad}
\label{eq:benchmark_dephasing_sec3_new}
\end{equation}
toward the upper end of the valid waveform band.

These numbers define a clear observational hierarchy. First, both the baseline and transition branches remain detectable at the level of the present illustrative LISA setup. Second, the mismatch is small, which means that after maximization over $t_c$ and $\phi_c$ the transition waveform remains highly overlapping with the baseline branch. Third, the residual norm is nonetheless of order unity, indicating that the structured difference between the two branches is not negligible in the detector-weighted norm. Finally, the accumulated phase drift is extremely large, showing that the transition perturbs the inspiral clock coherently over a long interval.

Taken together, these results place the benchmark in a bias-sensitive regime:
\begin{equation}
\mathcal M\ll1,
\qquad
\rho_{\rm R}\sim1,
\qquad
\Delta\Phi_{22}\gg1.
\label{eq:hierarchy_sec3_new}
\end{equation}
The main implication is therefore not that the signal becomes undetectable, but that a baseline analysis may remain observationally adequate for recovery while becoming insufficiently faithful for precision inference. This hierarchy is the organizing principle for the numerical results presented in the next section.

\section{Results and Discussion}
\label{sec:results}

We now present the quantitative consequences of the transition-modified Kerr EMRI model introduced in Sec.~\ref{sec:model} and interpreted through the LISA data-analysis framework of Sec.~\ref{sec:data_analysis}. The central result of this section is that the benchmark waveform simultaneously exhibits three properties: it remains detectable, it develops a very large cumulative phase drift, and it leaves a structured residual that is already large enough to matter for precision inference.

For the benchmark configuration
\begin{equation}
M=2.0\times10^5M_\odot,\qquad
\mu=1.4M_\odot,\qquad
\hat a=0.90,
\label{eq:benchmark_results_A}
\end{equation}
with the mixed-phase window
\begin{equation}
f_{22}\in[0.012,\,0.021]~{\rm Hz},
\label{eq:window_results_A}
\end{equation}
the principal detector-weighted diagnostics are
\begin{align}
\rho_{\rm B} &= 5.064,\\
\rho_{\rm T} &= 4.073,\\
\rho_{\rm R} &= 1.051,\\
\mathcal M &= 2.986\times10^{-3},
\label{eq:benchmark_metrics_A}
\end{align}
while the accumulated phase difference reaches
\begin{equation}
\Delta\Phi_{22}^{\rm SF}\sim 5\times10^3~{\rm rad}
\label{eq:benchmark_dphi_A}
\end{equation}
before the waveform is terminated at the flux-safe cutoff. These quantities must be interpreted jointly rather than in isolation.

\begin{table*}[t]
\centering
\caption{Benchmark source parameters, transition-sector setup, and detector-weighted diagnostics for the fiducial transition-modified Kerr EMRI.}
\label{tab:benchmark_summary}
\begin{tabular}{lll}
\hline
Category & Quantity & Value \\
\hline
Source parameters
& $M$ & $2.0\times10^5\,M_\odot$ \\
& $\mu$ & $1.4\,M_\odot$ \\
& $\hat a$ & $0.90$ \\
\hline
Transition sector
& mixed-phase window & $f_{22}\in[0.012,\,0.021]~{\rm Hz}$ \\
& flux-safe cutoff & $f_{22}^{\rm cut}\simeq 0.0364~{\rm Hz}$ \\
& pre-transition branch & $\Lambda_{\rm H}\sim 450$ \\
& post-transition branch & $\Lambda_{\rm Q}\ll \Lambda_{\rm H}$ \\
\hline
Detector-weighted metrics
& baseline SNR & $\rho_{\rm B}=5.064$ \\
& transition SNR & $\rho_{\rm T}=4.073$ \\
& residual SNR & $\rho_{\rm R}=1.051$ \\
& mismatch & $\mathcal M=2.986\times10^{-3}$ \\
& accumulated phase drift & $\Delta\Phi_{22}^{\rm SF}\sim 5\times10^3~{\rm rad}$ \\
\hline
Derived diagnostics
& fractional SNR suppression & $1-\rho_{\rm T}/\rho_{\rm B}\approx 0.196$ \\
& residual-to-baseline ratio & $\rho_{\rm R}/\rho_{\rm B}\approx 0.208$ \\
& accumulated phase in cycles & $\Delta\Phi_{22}^{\rm SF}/(2\pi)\sim 8\times10^2$ \\
\hline
\end{tabular}
\end{table*}

\subsection{Benchmark hierarchy and global waveform imprint}

The benchmark already defines a clear hierarchy,
\begin{equation}
\mathcal M\ll1,
\qquad
\rho_{\rm R}\sim1,
\qquad
\Delta\Phi_{22}^{\rm SF}\gg1,
\label{eq:hierarchy_A}
\end{equation}
which captures the main operational conclusion of the paper: the first-order transition is not primarily a missed-detection problem, but a precision-inference problem.

\begin{figure*}[t]
\centering
\includegraphics[width=0.5\textwidth]{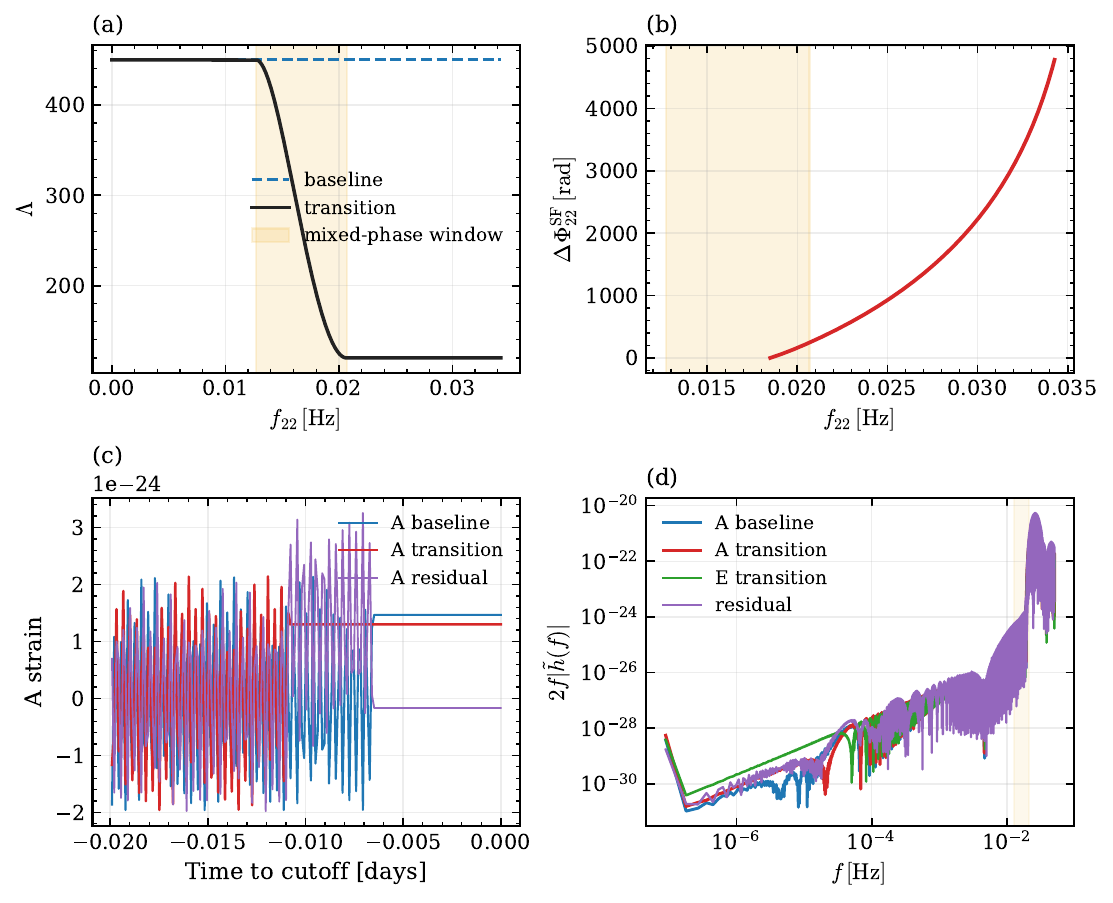}
\caption{
Global summary of the first-order phase-transition imprint on the Kerr EMRI waveform.
(a) Effective transition-control parameter $\Lambda$ as a function of the dominant-mode frequency $f_{22}$, with the mixed-phase interval highlighted.
(b) Accumulated phase difference $\Delta\Phi_{22}^{\rm SF}$ between the baseline and transition branches.
(c) Time-domain residual structure near the end of the inspiral.
(d) Frequency-domain characteristic strains of the baseline, transition, and residual branches relative to the LISA sensitivity scale.
The figure shows that a localized restructuring in the dissipative sector can generate a large coherent phase defect while the two main branches remain globally close in detector-weighted norm.
}
\label{fig:main4panel}
\end{figure*}

Figure~\ref{fig:main4panel} provides the global summary of the transition imprint. Panel (a) shows the effective transition-control parameter $\Lambda$ across the mixed-phase interval. Panel (b) shows the accumulated phase difference $\Delta\Phi_{22}^{\rm SF}$, which grows coherently toward the upper end of the valid waveform band. Panels (c) and (d) connect this phase reorganization to the time-domain residual and the frequency-domain characteristic-strain structure, respectively.

\begin{figure}[t]
\centering

\includegraphics[width=0.5\textwidth]{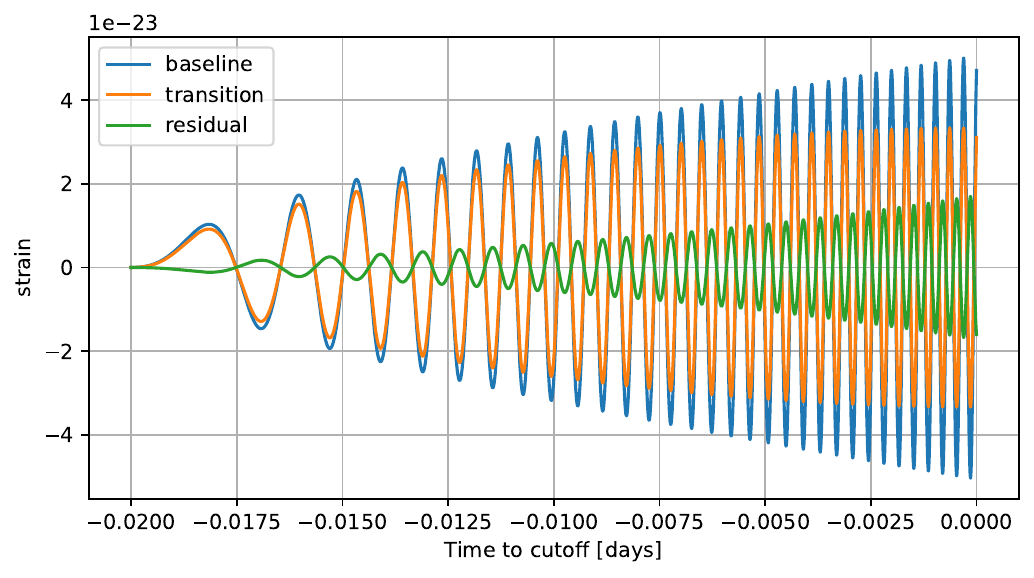}
\caption{
Time-domain comparison of the baseline, transition, and residual strains as functions of time to cutoff.
The baseline and transition waveforms remain visually close at early times but develop increasing phase slippage toward the end of the inspiral.
The residual is therefore generated primarily by coherent dephasing rather than by a large amplitude mismatch.
}
\label{fig:timeoverlay}
\end{figure}

The time-domain interpretation is shown separately in Fig.~\ref{fig:timeoverlay}. The baseline and transition strains remain close at earlier times but progressively lose phase alignment toward the end of the inspiral. The residual is therefore generated primarily by coherent phase slippage rather than by an isolated amplitude feature.

This structure can be understood by writing
\begin{align}
h_{\rm B}(t) &= A(t)\cos\Phi_{\rm B}(t),\\
h_{\rm T}(t) &= \tilde A(t)\cos\Phi_{\rm T}(t),
\end{align}
with $\tilde A(t)\approx A(t)$ and $\delta\Phi(t)\equiv\Phi_{\rm T}(t)-\Phi_{\rm B}(t)$. The residual then takes the approximate form
\begin{equation}
h_{\rm R}(t)
\approx
2A(t)\sin\!\left(\frac{\delta\Phi(t)}{2}\right)
\sin\!\left(\frac{\Phi_{\rm B}(t)+\Phi_{\rm T}(t)}{2}\right),
\label{eq:residual_phase_form_A}
\end{equation}
showing directly that the remainder is controlled by the accumulated phase drift.

\begin{figure}[t]
\centering
\includegraphics[width=0.5\textwidth]{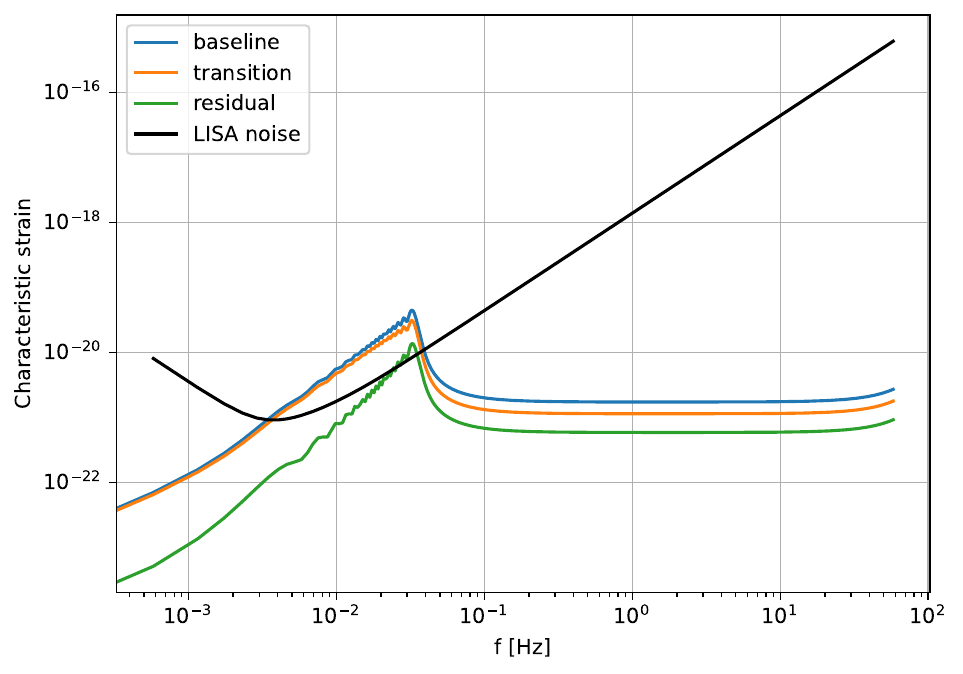}
\caption{
Characteristic strain of the baseline, transition, and residual waveforms compared with the LISA noise level.
The baseline and transition branches remain close over most of the analyzed band, explaining the small mismatch, while the residual retains visible support near the upper inspiral band where the accumulated phase defect becomes most important.
}
\label{fig:charstrain}
\end{figure}

The frequency-domain interpretation is shown in Fig.~\ref{fig:charstrain}. The baseline and transition characteristic strains remain close across most of the analyzed band, explaining the small mismatch, while the residual remains visibly supported near the upper inspiral band where the accumulated phase defect becomes most consequential.

Taken together, Figs.~\ref{fig:main4panel}--\ref{fig:charstrain} show why a small mismatch and a large coherent phase displacement are not contradictory. The mismatch is a normalized detector-weighted projection, whereas $\Delta\Phi_{22}^{\rm SF}$ is a cumulative dynamical observable. In a long-lived EMRI, the latter is the more serious warning sign for faithful inference.

\subsection{Residual significance and bias susceptibility}

The decisive scalar measure of the unmodeled remainder is the residual SNR,
\begin{equation}
\rho_{\rm R} = \sqrt{(h_{\rm R}|h_{\rm R})}=1.051,
\label{eq:rhoR_A}
\end{equation}
which places the benchmark near the practical distinguishability threshold
\begin{equation}
(\delta h|\delta h)^{1/2}\sim1.
\label{eq:threshold_A}
\end{equation}
Thus, the residual is not large enough to define a completely separate high-SNR signal class, but it is large enough that it cannot be neglected in precision waveform modeling.

At the same time, the mismatch remains very small,
\begin{equation}
\mathcal M\simeq 2.986\times10^{-3},
\label{eq:mismatch_A}
\end{equation}
which means that after maximization over the extrinsic offsets $t_c$ and $\phi_c$, the transition waveform remains highly overlapping with the baseline branch. This is sufficient for source recovery in the restricted sense actually evaluated here, but it is not sufficient to guarantee faithful intrinsic inference.

\begin{figure}[t]
\centering
% NEW FIGURE 1
\includegraphics[width=0.5\textwidth]{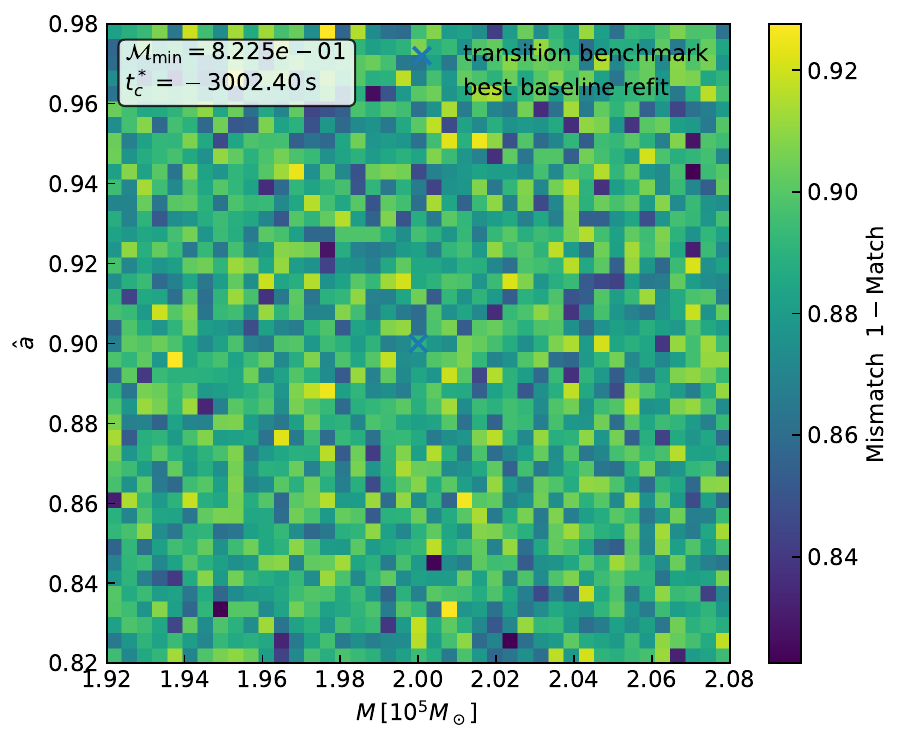}
\caption{
Detector-weighted mismatch after local intrinsic refitting of the baseline branch in a representative parameter plane, use $(M,\hat a)$ .
The benchmark point and the best-fit intrinsic baseline point should both be indicated.
This figure is intended to test how effectively the transition waveform can be mimicked by nearby Kerr parameters.
}
\label{fig:refit_mismatch}
\end{figure}

To make this point quantitative, Fig.~\ref{fig:refit_mismatch} should show the detector-weighted mismatch after local intrinsic refitting of the baseline branch in a representative parameter plane such as $(M,\hat a)$ or $(M,\mu)$. This figure tests whether the transition waveform can be reabsorbed by nearby intrinsic points of the baseline family.

The inferential geometry is then captured by the Fisher-bias formula
\begin{equation}
\Delta\theta_{\rm sys}^i
=
(\Gamma^{-1})^{ij}
\left(\delta h\middle|\partial_j h_{\rm B}\right),
\label{eq:fisher_bias_A}
\end{equation}
with
\begin{equation}
\Gamma_{ij}
=
\left(\partial_i h_{\rm B}\middle|\partial_j h_{\rm B}\right).
\end{equation}
The most vulnerable directions are expected to be those controlling phase accumulation, especially
\begin{equation}
\theta^i\in\{M,\mu,\hat a\},
\end{equation}
with additional correlations involving $t_c$ and $\phi_c$.

\begin{figure}[t]
\centering
% NEW FIGURE 2
\includegraphics[width=0.5\textwidth]{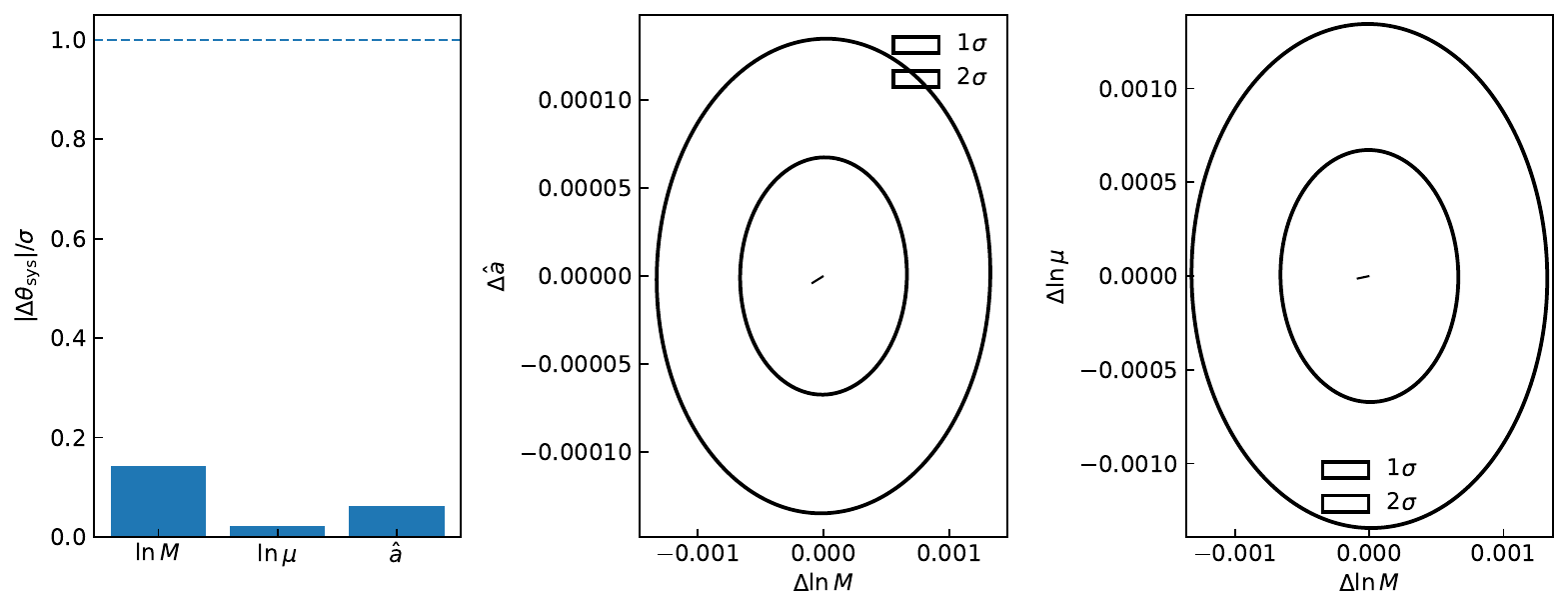}
\caption{
Fisher-bias diagnostic for the transition benchmark.
Suggested implementations are either:
(i) a bar chart of $|\Delta\theta_{\rm sys}|/\sigma$ for the dominant parameters, or
(ii) Fisher confidence ellipses with the systematic-shift vector overlaid.
This figure quantifies the extent to which the missing transition sector biases precision inference even when the normalized mismatch remains small.
}
\label{fig:fisher_bias}
\end{figure}

Figure~\ref{fig:fisher_bias} should therefore display either the bias significance $|\Delta\theta_{\rm sys}|/\sigma$ for the dominant parameters or Fisher confidence ellipses with the systematic-shift vector overlaid.

The physical interpretation is straightforward: if the waveform family lacks an explicit transition sector, the missing phase physics must be reabsorbed into biased estimates of ordinary Kerr parameters. This is the precise sense in which the benchmark lies in a bias-sensitive regime.

\subsection{Robustness beyond the benchmark}

The benchmark point establishes the existence of a bias-prone regime, but by itself it does not show how broad that regime is. For this reason, the results section should also include a two-dimensional scan over the mixed-phase window, for example in the window center $f_c$ and width $\Delta f$.

\begin{figure*}[t]
\centering
\includegraphics[width=0.98\textwidth]{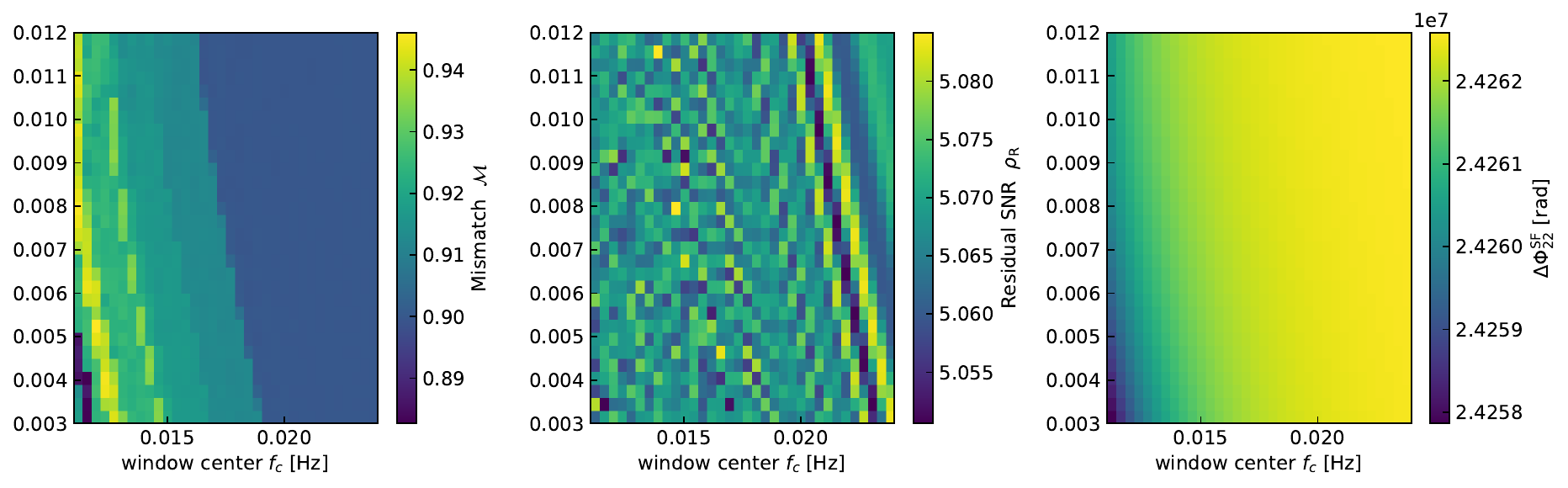}
\caption{
Robustness of the transition imprint across the mixed-phase parameter space.
Suggested implementation: a two-dimensional scan over the window center $f_c$ and width $\Delta f$, with color maps for the mismatch $\mathcal M$, residual SNR $\rho_{\rm R}$, and accumulated phase drift $\Delta\Phi_{22}^{\rm SF}$.
This figure should determine whether the benchmark is an isolated point or part of a broader region in which small mismatch coexists with large dephasing and order-unity residual norm.
}
\label{fig:window_scan}
\end{figure*}

Figure~\ref{fig:window_scan} should show color maps for the mismatch $\mathcal M$, the residual SNR $\rho_{\rm R}$, and the accumulated phase drift $\Delta\Phi_{22}^{\rm SF}$. Its purpose is to determine whether the benchmark is an isolated point or part of a broader region in which small mismatch coexists with large dephasing and order-unity residual norm.

If the hierarchy
\begin{equation}
\mathcal M\ll1,\qquad
\rho_{\rm R}\sim1,\qquad
\Delta\Phi_{22}^{\rm SF}\gg1
\end{equation}
persists over a finite region of transition-parameter space, then the present benchmark is not a numerical curiosity but a representative example of a broader inference problem. The implication for future EMRI waveform development is immediate: the waveform family should be extended to parameterize at least the location, width, and amplitude of the mixed-phase correction, together with the pre-transition and post-transition branch structure. Without such an extension, the data-analysis pipeline has only one option: to absorb missing transition physics into biased estimates of ordinary source parameters.

\section{Conclusion}
\label{sec:conclusion}

In this work, we constructed a flux-safe phenomenological framework to study how a first-order phase-transition (FOPT) sector can imprint itself on Kerr extreme mass-ratio inspiral (EMRI) waveforms in the LISA band. The analysis combined three ingredients: an adiabatic circular equatorial inspiral model in Kerr spacetime, a finite-width transition sector acting in the dissipative fluxes, and a detector-weighted interpretation based on standard matched filtering. Within this setup, the transition does not enter as a catastrophic disruption of the waveform, but as a localized restructuring of the inspiral clock whose effects accumulate coherently over the subsequent evolution.

The main physical conclusion is that a narrow transition in the effective flux sector can generate a very large cumulative phase displacement even when the transition waveform remains highly overlapping with the baseline Kerr branch after maximization over the extrinsic shifts. For the benchmark configuration
\begin{equation*}
M=2.0\times10^5M_\odot,\qquad
\mu=1.4M_\odot,\qquad
\hat a=0.90,
\end{equation*}
with the mixed-phase window placed in
\begin{equation*}
f_{22}\in[0.012,\,0.021]~{\rm Hz},
\end{equation*}
we obtained a small mismatch,
\begin{equation*}
\mathcal M \simeq 2.986\times10^{-3},
\end{equation*}
together with a large accumulated phase drift,
\begin{equation*}
\Delta\Phi_{22}^{\rm SF}\sim 5\times10^3~{\rm rad},
\end{equation*}
and a residual norm of order unity,
\begin{equation*}
\rho_{\rm R}\simeq 1.051.
\end{equation*}
These quantities define a specific observational hierarchy:
\begin{equation*}
\mathcal M\ll 1,
\qquad
\rho_{\rm R}\sim 1,
\qquad
\Delta\Phi_{22}^{\rm SF}\gg 1.
\end{equation*}
The waveform therefore remains observationally close to the baseline branch in normalized detector space, while the unmodeled remainder and the cumulative phase defect are already too large to be ignored in precision interpretation.

This separation between detectability and faithfulness is the central message of the paper. In an EMRI, a transition need not occupy a large fraction of the observed band in order to matter. It is sufficient that the transition perturb the radiation-reaction law over a finite interval and thereby alter the time-frequency map of the inspiral. Once this happens, the later waveform carries an integrated dynamical memory of having crossed the mixed-phase region. The phase defect is therefore not a local glitch confined to the transition window, but a cumulative reorganization of the subsequent signal.

From the perspective of data analysis, the present benchmark is not primarily a missed-detection problem. Rather, it identifies a bias-sensitive regime in which a baseline Kerr analysis can remain adequate for source recovery in the restricted sense tested here, while becoming insufficiently faithful for precision inference. If the waveform family does not contain an explicit transition sector, the missing phase physics will tend to project onto ordinary source parameters such as the masses, the spin, and the phase-evolution constants. In this sense, the dominant consequence of the FOPT sector is not loss of detectability, but susceptibility to systematic bias.

The broader implication is methodological. Precision EMRI astronomy may ultimately require waveform families that go beyond purely vacuum Kerr inspirals, even when non-vacuum effects are phenomenologically narrow. A finite transition sector can be narrow in frequency space yet broad in inference space. Future waveform development should therefore move toward parameterized non-vacuum Kerr EMRI models in which the location, width, and amplitude of the mixed-phase correction are incorporated directly into the waveform manifold. Only then can one ask, in a statistically controlled way, whether the data prefer a purely Kerr inspiral or a Kerr inspiral supplemented by a finite transition sector.

The present study should be regarded as a controlled first step rather than a final waveform model. Several extensions are immediate: replacing the effective PN-like flux by Teukolsky-based or self-force-informed fluxes, including harmonic content beyond the dominant $(2,2)$ mode, quantifying local intrinsic refit and Fisher-bias diagnostics in the full parameter space, and performing Bayesian model comparison between transition and non-transition waveform families. These developments will determine how robustly the effect identified here survives in more realistic LISA analyses.

Nevertheless, the central conceptual result is already clear. In a Kerr EMRI, a first-order phase transition acting through the inspiral fluxes can leave a waveform that is globally similar to the standard baseline signal yet cumulatively different enough to compromise precision interpretation. The source may still be found, but it is not fully understood unless the transition sector is modeled. This is precisely the type of effect that becomes important in the era of high-accuracy space-based gravitational-wave astronomy.

\appendix

\section{Technical details of the inspiral model}
\label{app:technical_model}

This appendix collects several technical relations underlying the phenomenological transition-modified Kerr EMRI model used in the main text. The goal is not to introduce a new physical sector beyond Sec.~\ref{sec:model}, but to record the explicit formulas most useful for implementation and reproducibility.

\subsection{Circular equatorial Kerr kinematics}

For a circular equatorial orbit in Kerr spacetime, the orbital angular velocity measured at infinity is
\begin{equation}
\Omega(r,\hat a)
=
\frac{1}{M\left[(r/M)^{3/2}+\hat a\right]},
\label{eq:Omega_app}
\end{equation}
where $\hat a=a/M$ is the dimensionless Kerr spin parameter. Introducing the invariant PN-like velocity variable
\begin{equation}
v\equiv (M\Omega)^{1/3},
\label{eq:v_def_app}
\end{equation}
the inverse-radius variable may be written as
\begin{equation}
u(v,\hat a)
\equiv
\frac{M}{r}
=
\frac{v^2}{\left(1-\hat a v^3\right)^{2/3}}.
\label{eq:u_of_v_app}
\end{equation}
The specific orbital energy of the small body is then
\begin{equation}
\tilde E(v,\hat a)
=
\frac{1-2u+\hat a u^{3/2}}
{\sqrt{1-3u+2\hat a u^{3/2}}},
\qquad
u=u(v,\hat a),
\label{eq:Etilde_app}
\end{equation}
so that the orbital energy is
\begin{equation}
E_{\rm orb}(v)=\mu\,\tilde E(v,\hat a).
\label{eq:Eorb_app}
\end{equation}

To construct the inspiral clock, one needs the derivative
\begin{equation}
\frac{d\tilde E}{dv}
=
\frac{d\tilde E}{du}\frac{du}{dv},
\label{eq:dEdv_chain_app}
\end{equation}
with
\begin{equation}
\frac{du}{dv}
=
\frac{2v}{\left(1-\hat a v^3\right)^{5/3}},
\label{eq:dudv_app}
\end{equation}
and
\begin{equation}
\frac{d\tilde E}{du}
=
\frac{2N'(u)D(u)-N(u)D'(u)}
{2D(u)^{3/2}},
\label{eq:dEdu_app}
\end{equation}
where
\begin{align}
N(u) &= 1-2u+\hat a u^{3/2}, \\
D(u) &= 1-3u+2\hat a u^{3/2}, \\
N'(u) &= -2+\frac{3}{2}\hat a u^{1/2}, \\
D'(u) &= -3+3\hat a u^{1/2}.
\end{align}

\subsection{Baseline and transition-modified inspiral clocks}

The inspiral is driven by the balance equation
\begin{equation}
\frac{dE_{\rm orb}}{dt}
=
-\mathcal F(v,\hat a).
\label{eq:balance_app}
\end{equation}
In the baseline model,
\begin{equation}
\mathcal F_{\rm B}(v,\hat a)
=
\frac{32}{5}\left(\frac{\mu}{M}\right)^2 v^{10}\,
\hat{\mathcal F}_{\rm PP}(v,\hat a),
\label{eq:Fbaseline_app}
\end{equation}
with the low-order PN-like effective Kerr flux factor
\begin{equation}
\hat{\mathcal F}_{\rm PP}(v,\hat a)
=
1-\frac{1247}{336}v^2
+\left(4\pi-\frac{73}{12}\hat a\right)v^3
-\frac{44711}{9072}v^4
+\mathcal O(v^5).
\label{eq:Fhat_pp_app}
\end{equation}
The corresponding inspiral clock is
\begin{equation}
\frac{dt_{\rm B}}{dv}
=
-\mu\frac{d\tilde E/dv}{\mathcal F_{\rm B}(v,\hat a)}.
\label{eq:dtdv_baseline_app}
\end{equation}

The phenomenological transition sector is introduced through the effective coupling
\begin{equation}
\Lambda(v)
=
\Lambda_{\rm H}
+
\left(\Lambda_{\rm Q}-\Lambda_{\rm H}\right)
S\!\left(\frac{v-v_{\rm h}}{v_{\rm q}-v_{\rm h}}\right),
\label{eq:Lambda_app}
\end{equation}
where the interpolation function is the $C^1$ smoothstep
\begin{equation}
S(x)=
\begin{cases}
0, & x<0,\\[4pt]
3x^2-2x^3, & 0\le x\le 1,\\[4pt]
1, & x>1.
\end{cases}
\label{eq:smoothstep_app}
\end{equation}
The transition-corrected flux is written as
\begin{equation}
\mathcal F_{\rm T}(v,\hat a)
=
\mathcal F_{\rm B}(v,\hat a)
\left[
1+\Lambda(v)\,\mathcal H_{\rm PT}(v,\hat a)
\right],
\label{eq:Ftransition_app}
\end{equation}
with response kernel
\begin{equation}
\mathcal H_{\rm PT}(v,\hat a)
=
\kappa_0 v^{10}
\left(
1+h_2 v^2+h_{3s}\hat a v^3+h_4 v^4
\right).
\label{eq:Hpt_app}
\end{equation}
The transition-driven clock is therefore
\begin{equation}
\frac{dt_{\rm T}}{dv}
=
-\mu\frac{d\tilde E/dv}{\mathcal F_{\rm T}(v,\hat a)}.
\label{eq:dtdv_transition_app}
\end{equation}

For the phenomenological evolution to remain well defined, we require
\begin{equation}
\hat{\mathcal F}_{\rm PP}(v,\hat a)>0,
\qquad
1+\Lambda(v)\mathcal H_{\rm PT}(v,\hat a)>0
\label{eq:positivity_app}
\end{equation}
throughout the integration interval. In practice, the waveform is terminated before the PN-like baseline flux develops its artificial zero, which leads to the flux-safe cutoff used in the main text,
\begin{equation}
f_{22}^{\rm cut}\simeq 0.0364~{\rm Hz}.
\label{eq:fcut_app}
\end{equation}
This cutoff is a model-validity boundary and should not be interpreted as the physical plunge frequency.  

\subsection{Phase accumulation and residual waveform}

The dominant $(2,2)$-mode phase obeys
\begin{equation}
\frac{d\Phi_{22}}{dt}=2\Omega(v,\hat a),
\label{eq:dphidt_app}
\end{equation}
and hence
\begin{equation}
\frac{d\Phi_{22}}{dv}
=
2\Omega(v,\hat a)\frac{dt}{dv}.
\label{eq:dphidv_app}
\end{equation}
For the baseline and transition branches,
\begin{align}
\frac{d\Phi_{22}^{\rm B}}{dv}
&=
2\Omega(v,\hat a)\frac{dt_{\rm B}}{dv},
\\
\frac{d\Phi_{22}^{\rm T}}{dv}
&=
2\Omega(v,\hat a)\frac{dt_{\rm T}}{dv},
\end{align}
so that the accumulated phase difference is
\begin{equation}
\Delta\Phi_{22}^{\rm SF}(v)
=
\Phi_{22}^{\rm B}(v)-\Phi_{22}^{\rm T}(v)
=
2\int^v \Omega(v')
\left[
\frac{dt_{\rm B}}{dv'}-\frac{dt_{\rm T}}{dv'}
\right]dv'.
\label{eq:DeltaPhi_int_app}
\end{equation}
This expression makes the central mechanism explicit: the transition first perturbs the inspiral clock and the phase then integrates that perturbation over the remaining evolution, allowing even a narrow mixed-phase interval to produce a macroscopically large dephasing. This is the same physical mechanism emphasized in the main text. :contentReference[oaicite:1]{index=1}

For a weak relative correction,
\begin{equation}
\mathcal F_{\rm T}(v)=\mathcal F_{\rm B}(v)\,[1+\epsilon(v)],
\qquad
|\epsilon(v)|\ll1,
\label{eq:eps_app}
\end{equation}
one has to first order
\begin{equation}
\frac{dt_{\rm T}}{dv}
\approx
\frac{dt_{\rm B}}{dv}\,[1-\epsilon(v)],
\label{eq:dtdv_weak_app}
\end{equation}
and therefore
\begin{equation}
\Delta\Phi_{22}^{\rm SF}(v)
\approx
2\int^v \Omega(v')\,
\frac{dt_{\rm B}}{dv'}\,
\epsilon(v')\,dv'.
\label{eq:DeltaPhi_weak_app}
\end{equation}

The dominant mode in the time domain is written as
\begin{equation}
h_{22}(t)
=
A_{22}(t)\cos\!\big[\Phi_{22}(t)+\phi_0\big],
\label{eq:h22_time_app}
\end{equation}
with leading-order amplitude
\begin{equation}
A_{22}(t)
=
\frac{2\mu}{D_L}\,v(t)^2\,\mathcal A_{22}(v,\hat a,\Lambda),
\label{eq:A22_app}
\end{equation}
where $D_L$ is the luminosity distance. The baseline, transition, and residual strains are
\begin{align}
h_{\rm B}(t) &= A_{\rm B}(t)\cos\Phi_{\rm B}(t),\\
h_{\rm T}(t) &= A_{\rm T}(t)\cos\Phi_{\rm T}(t),\\
h_{\rm R}(t) &= h_{\rm B}(t)-h_{\rm T}(t).
\label{eq:residual_app}
\end{align}
When amplitude differences are subleading, the residual may be approximated by
\begin{equation}
h_{\rm R}(t)
\approx
2A(t)\sin\!\left(\frac{\delta\Phi(t)}{2}\right)
\sin\!\left(\frac{\Phi_{\rm B}(t)+\Phi_{\rm T}(t)}{2}\right),
\label{eq:residual_phase_form_app}
\end{equation}
where $\delta\Phi(t)\equiv\Phi_{\rm T}(t)-\Phi_{\rm B}(t)$. This clarifies why the residual is controlled primarily by coherent phase slippage rather than by a large amplitude burst. The same qualitative point is made in the main results section using the time-domain comparison plot. :contentReference[oaicite:2]{index=2}

\section{Detector conventions and inference diagnostics}
\label{app:data_analysis}

This appendix records the detector-weighted conventions used in the matched-filter analysis and the corresponding benchmark hierarchy.

\subsection{Frequency-domain representation and inner product}

The Fourier transform convention is
\begin{equation}
\tilde h(f)=\int_{-\infty}^{+\infty} h(t)\,e^{-2\pi i f t}\,dt.
\label{eq:FT_app}
\end{equation}
The noise-weighted inner product between two waveforms is
\begin{equation}
(h_1|h_2)
=
4\,\mathrm{Re}\int_{f_{\min}}^{f_{\max}}
\frac{\tilde h_1(f)\tilde h_2^*(f)}{S_n(f)}\,df,
\label{eq:inner_product_app}
\end{equation}
where $S_n(f)$ is the one-sided LISA noise power spectral density. The integration limits are chosen to match the modeled signal band,
\begin{equation}
f_{\min}=f_{22}^{\rm start},
\qquad
f_{\max}=f_{22}^{\rm cut}.
\label{eq:band_app}
\end{equation}

A standard sky-averaged analytic representation of the instrumental contribution may be written schematically as
\begin{equation}
S_{\rm inst}(f)
=
\frac{10}{3L^2}
\left[
P_{\rm OMS}(f)+\frac{4P_{\rm acc}(f)}{(2\pi f)^4}
\right]
\left[
1+\frac{6}{10}\left(\frac{f}{f_*}\right)^2
\right],
\label{eq:Sinst_app}
\end{equation}
with $f_*=c/(2\pi L)$. In the main text this detector model is used consistently across the baseline, transition, and residual branches so that all comparisons remain detector weighted. 

\subsection{SNR, overlap, mismatch, and residual norm}

The optimal matched-filter signal-to-noise ratio is
\begin{equation}
\rho[h]=\sqrt{(h|h)}.
\label{eq:snr_app}
\end{equation}
For the three waveform branches,
\begin{align}
\rho_{\rm B} &= \sqrt{(h_{\rm B}|h_{\rm B})},\\
\rho_{\rm T} &= \sqrt{(h_{\rm T}|h_{\rm T})},\\
\rho_{\rm R} &= \sqrt{(h_{\rm R}|h_{\rm R})}.
\label{eq:snr_branches_app}
\end{align}
For the benchmark system used throughout the paper, the numerical values are
\begin{equation}
\rho_{\rm B}=5.064,
\qquad
\rho_{\rm T}=4.073,
\qquad
\rho_{\rm R}=1.051.
\label{eq:snr_values_app}
\end{equation}
These are the same benchmark numbers quoted in the revised results and conclusion. 

The normalized overlap is
\begin{equation}
\mathcal O(h_1,h_2)
=
\frac{(h_1|h_2)}
{\sqrt{(h_1|h_1)(h_2|h_2)}},
\label{eq:overlap_app}
\end{equation}
and the match is defined by maximizing the overlap over the extrinsic offsets $t_c$ and $\phi_c$,
\begin{equation}
\mathrm{Match}
=
\max_{t_c,\phi_c}
\mathcal O\!\left(
h_{\rm B},\,
h_{\rm T}e^{i(2\pi f t_c+\phi_c)}
\right).
\label{eq:match_app}
\end{equation}
The mismatch is then
\begin{equation}
\mathcal M=1-\mathrm{Match}.
\label{eq:mismatch_app}
\end{equation}
For the benchmark model,
\begin{equation}
\mathcal M=2.986\times10^{-3}.
\label{eq:mismatch_value_app}
\end{equation}
This number is small, but it should be interpreted as a detector-weighted proximity measure under extrinsic alignment rather than as a complete intrinsic-parameter search. This distinction is central to the main paper’s interpretation. 

The residual waveform is
\begin{equation}
\delta h \equiv h_{\rm T}-h_{\rm B}=-h_{\rm R},
\label{eq:deltah_app}
\end{equation}
and its norm is
\begin{equation}
(\delta h|\delta h)^{1/2}=\rho_{\rm R}.
\label{eq:deltah_norm_app}
\end{equation}
A commonly used distinguishability criterion is
\begin{equation}
(\delta h|\delta h)^{1/2}\gtrsim 1.
\label{eq:distinguishability_app}
\end{equation}
The benchmark value $\rho_{\rm R}\simeq1.051$ therefore places the model near the threshold where the unmodeled remainder becomes observationally relevant after subtraction of the baseline waveform. :contentReference[oaicite:6]{index=6}

For spectral visualization, we also use the characteristic strain
\begin{equation}
h_c(f)=2f\,|\tilde h(f)|,
\label{eq:hc_app}
\end{equation}
and the corresponding detector reference scale
\begin{equation}
h_n(f)=\sqrt{fS_n(f)}.
\label{eq:hn_app}
\end{equation}
These are the quantities plotted in the characteristic-strain figure in the main text. :contentReference[oaicite:7]{index=7}

\subsection{Fisher-bias framework and benchmark hierarchy}

The local geometry of the baseline waveform manifold is encoded in the Fisher matrix
\begin{equation}
\Gamma_{ij}
=
\left(\partial_i h_{\rm B}\middle|\partial_j h_{\rm B}\right),
\label{eq:Fisher_app}
\end{equation}
and, if the true signal is
\begin{equation}
h_{\rm T}=h_{\rm B}+\delta h,
\end{equation}
the leading-order systematic shift is
\begin{equation}
\Delta\theta_{\rm sys}^i
=
(\Gamma^{-1})^{ij}
\left(\delta h\middle|\partial_j h_{\rm B}\right).
\label{eq:bias_app}
\end{equation}
This is the precise sense in which a small mismatch does not guarantee faithful recovery: even when $\delta h$ is not large enough to prevent detection, it can still project strongly onto the tangent directions of the baseline model. The main paper identifies the most vulnerable directions as those controlling phase accumulation, especially $M$, $\mu$, and $\hat a$. 

The combined benchmark diagnostics define the observational hierarchy
\begin{equation}
\mathcal M\ll 1,
\qquad
\rho_{\rm R}\sim 1,
\qquad
\Delta\Phi_{22}^{\rm SF}\gg 1.
\label{eq:hierarchy_app}
\end{equation}
For the explicit benchmark,
\begin{equation}
\mathcal M\simeq 2.986\times10^{-3},
\qquad
\rho_{\rm R}\simeq1.051,
\qquad
\Delta\Phi_{22}^{\rm SF}\sim5\times10^3~{\rm rad},
\label{eq:hierarchy_benchmark_app}
\end{equation}
which is the core numerical hierarchy used throughout the revised results and conclusion. 

\section{Supplementary visual comparison of phase slippage}
\label{app:phase_slippage_visual}

For completeness, we include two supplementary time-domain visualizations of the phase slippage between the baseline and transition-modified chirps. These figures are qualitative only and are intended as visual complements to the quantitative diagnostics presented in the main text.

\begin{figure}[t]
\centering
\includegraphics[width=0.5\textwidth]{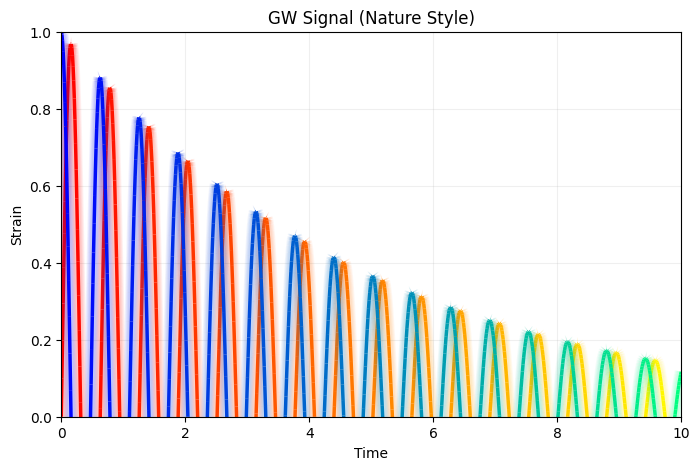}
\caption{
Supplementary visualization of cumulative phase slippage between the baseline and transition branches.
}
\label{fig:schematic1}
\end{figure}

\begin{figure}[t]
\centering
\includegraphics[width=0.5\textwidth]{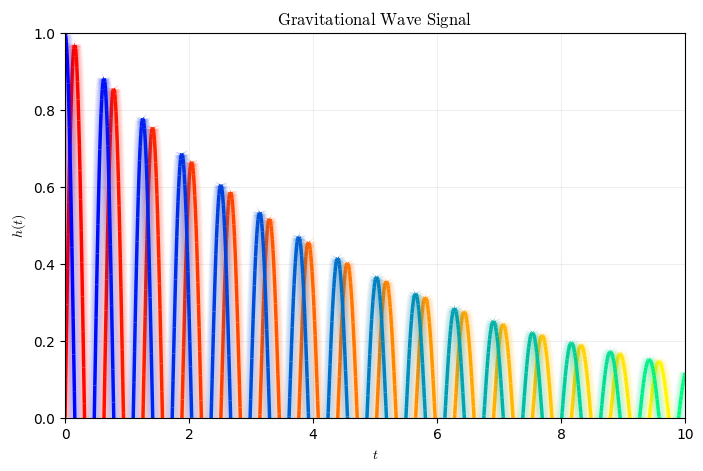}
\caption{
Alternative supplementary visualization of the same qualitative phase-slippage effect.
}
\label{fig:schematic2}
\end{figure}

\end{document}